# Avaliação da produtividade de hospitais brasileiros pela metodologia do *diagnosis related group* (DRG)

Evaluation of the productivity of Brazilian hospitals by the methodology of diagnosis related group (DRG)

La evaluación de la productividad de los hospitales brasileños por metodología de *diagnosis related group* (DRG)

*JOSÉ CARLOS SERUFO FILHO*

## Abstract

The management requires a hospital organization to provision their costs/expenses with tools that approximate reality. The task of measuring productivity can be complex and uncertain, several methods are tested and the use of the DRG has been efficient, being used to assess the productivity through clinical outcomes. Cross-sectional study evaluated 145.710 hospitalizations in the period 2012-2014, using the DRG methodology for measuring productivity from the median length of hospitalization. When we group all hospitalizations in clinical (37.6%) and surgical (62.4%), multiple analyzes could be made according to this criterion. The DRG as a tool for prediction of hospital days is an effective alternative, thereby contributing to the control of productivity that directly influences the costs of hospital expenses and product and service quality.

## Resumo

O Gerenciamento de uma organização hospitalar exige provisionar seus custos/gastos com ferramentas que a aproximam da realidade. A tarefa de aferição da produtividade pode ser complexa e duvidosa, diversos métodos são experimentados e a utilização do DRG tem se mostrado eficiente, sendo utilizado na avaliação da produtividade através de desfechos assistenciais. Estudo transversal, avaliou 145.710 internações, no período de 2012-2014, utilizando a metodologia do DRG para medição de sua produtividade a partir da mediana do tempo de internação. Ao agruparmos todas as internações em clínicos (37,6%) e cirúrgicos (62,4%), várias análises puderam ser feitas de acordo com esse critério. O DRG como ferramenta para predição de dias de internação é uma alternativa eficiente, colaborando assim para o controle da produtividade que influencia diretamente nos gastos e custos dos produtos hospitalares e qualidade dos serviços.

## Introdução

Os sistemas de saúde mundiais consomem recursos vultuosos. De acordo com a Organização Mundial de Saúde (OMS) e dados do Banco Mundial em 2012, 17,91% do produto interno bruto (PIB) dos Estados Unidos da América (EUA) foi investido em saúde, o que corresponde ao valor de aproximadamente 2.9 trilhões de dólares. Outros países como a Suíça (11,30%), Holanda (12,44%), Argentina (8,49%) e Colômbia (6,83%) também apresentam parcelas significativas do orçamento dos governos investidas na saúde. O percentual do PIB brasileiro destinado à saúde foi de 9,31%, aproximadamente 200 bilhões de dólares. De acordo com a Constituição Federal, os municípios são obrigados a destinar 15% do que arrecadam em ações de saúde. Para os governos estaduais, esse percentual é de 12%. No caso da União, o formato é um pouco diferente, pois leva em consideração a arrecadação do ano anterior para definir seus gastos.

Existe grande variação entre os países em relação ao percentual do PIB destinado à saúde, com valores entre menos de 5% até um pouco mais de 20%. São fatores determinantes desta variação: PIB per capita, estrutura demográfica, características epidemiológicas da população, grau de progresso tecnológico, variações da prática médica e características do sistema de saúde [1]. Para tão elevado investimento esperam-se resultados assistenciais que mantenham boa relação com o consumo de recursos sociais.

Desenvolvido por Fetter *et al.* em 1980 para o governo Norte-Americano, o DRG (*Diagnosis Related Groups*) constitui um sistema de classificação de pacientes internados em hospitais que atendem casos agudos, ou seja, aqueles em que a média de permanência do paciente não ultrapassa 30 dias [2].

O sistema de classificação busca relacionar os tipos de pacientes atendidos pelo hospital, com os recursos consumidos durante o período de internação, criando grupos de pacientes que sejam semelhantes em suas características clínicas e no seu consumo de recursos [3].

A utilização do DRG possibilitou a realização de comparações da assistência hospitalar e seus custos, tornando possível a análise de produtividade através da relação entre os resultados assistenciais e econômicos [4]. A efetividade mensurada pelo DRG em uma organização poderá ser comparada com outras organizações hospitalares, criando oportunidades de melhoria de desempenho [5].

Assim, o DRG permite comparar o desempenho hospitalar entre instituições. Essa qualidade da classificação, associada à facilidade na obtenção dos dados de hospitalização encontrados em resumos de saídas hospitalares, é o que permite sua utilização para diversos propósitos dentro da grande área de gestão dos serviços hospitalares. Sua aplicação tem sido voltada para o pagamento de hospitalizações e para o planejamento e gestão do sistema de atenção hospitalar em seus diversos níveis [6].

Este estudo mensurou a produtividade dos leitos de hospitais brasileiros e estabeleceu comparação com os de hospitais americanos.

## Materiais e Métodos

Trata-se de estudo transversal, que comparou a mediana de permanência por produto DRG no Brasil com os percentis de permanência dos mesmos produtos assistidos na rede de hospitais americanos que vendem serviços ao governo americano, sendo as variações encontradas denominadas produtividade do uso do leito para gerar produtos assistenciais. Foram incluídos 147.542 indivíduos, no período de 2012 a 2014, classificados em 424 DRG e excluídos 1.832 que apresentavam DRG com menos de 20 pacientes. O projeto foi aprovado no COEP sob número 34133814.5.0000.5149. Uma vez que o banco de dados utilizado não identifica pacientes, foi obtida a dispensa do TCLE.

A análise do estudo incluiu: Sexo, Idade, CID principal, Comorbidades e Procedimentos; Peso da complexidade assistencial e Tipo de internação (Cirúrgico ou Clínico); Tempo de permanência hospitalar e Produtividade. Os percentis do tempo de internação americano foram utilizados como medidas de referência (critério) para análise dos resultados apresentados. Foram utilizados os testes do Qui-Quadrado, Análise de Correlação de *Pearson* e Teste de Sinal, respeitando uma probabilidade de significância inferior a 5% (**p** < 0,05). Utilizou-se o pacote estatístico SAS (SAS Institute Inc., 1989), o software R Studio Version 0.98.978 – © 2009-2013 RStudio, Inc, Epi Info verão 7 e SPSS versão 17 – cada um dentro de suas limitações e facilidades.

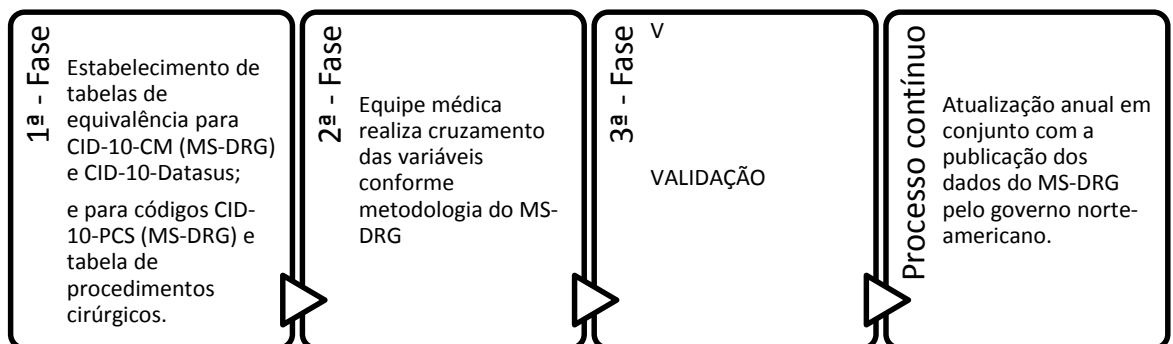
**Figura 1 - Fluxo do processo de ajuste do DRG-Brasil**

O CID-10, utilizado no Brasil, apresenta menor quantidade de códigos quando comparado ao utilizado pelo MS-DRG (CID-10-CM e CID-10-PCS) que categoriza as doenças com maior detalhamento. Portanto, fez-se necessário que uma equipe médica especializada realizasse a tarefa de verificação das similaridades, criando uma tabela de equivalências. A FIGURA 1 explica o fluxo de desenvolvimento e processo de criação e atualização do software DRG-Brasil.

Foi utilizada a categorização DRG do governo norte-americano em sua versão 31.0 (MS-DRG), cuja base CID-10-CM e CID-10-PCS. Para compatibilizar o sistema de codificação Brasileiro de procedimentos TUSS (Terminologia Unificada da Saúde Suplementar) e SUS (Sistema Único de Saúde) foi utilizado o software DRG Brasil® que correlaciona o sistema de códigos estadunidense com os códigos brasileiros. A validação das correspondências realizadas pelo software ocorreu produto a produto. O software utilizado no estudo foi desenvolvido e adaptado por médicos envolvidos em grupos de pesquisa da Universidade Federal de Minas Gerias e Faculdade de Ciências Médicas de Minas Gerais. Para o cálculo da produtividade foi utilizado como referência os dados publicados pelo governo norte-americano em sua rede de assistência para o MS-DRG, onde foi feita uma razão entre valores de permanência real nos hospitais estudados e o tempo de permanência estimado com base nos percentis americanos em cada um dos DRG avaliados. Assim, valores acima de 1,0 indicam o número de vezes que o tempo de internação nos hospitais estudados é maior que o previsto (hospitais americanos), ou seja, a produtividade é menor do que a obtida nos hospitais americanos. De forma análoga, números abaixo de 1,0 indicam ganho na produtividade.

## Resultados

O sexo feminino foi predominante (63,9%). A média da idade foi 42,8 anos e a mediana 39,9 anos ($P_{50}$). Um pouco mais de 60% dos pacientes (62,4%) foram classificados no DRG do grupo Cirúrgico e os demais 37,6% encontram-se inseridos no grupo Clínico. Os CID principais mais frequentes totalizaram 20.034 (36,6%) dos 54.808 (37,6%) dos casos de DRG clínicos, e 51.555 (56,7%) dos 90.902 (62,4%) dos casos de DRG cirúrgicos analisados.

Em relação à quantidade de diagnósticos múltiplos das subdivisões de DRG clínicos e cirúrgicos (TABELA 1), observa-se que no grupo de DRG clínicos 46,5% apresentaram pelo menos um diagnóstico secundário, seguidos por 22,0% que não apresentaram nenhum diagnóstico secundário. Enquanto no grupo de DRG cirúrgicos, quase da metade dos casos (47,4%) não tiveram relatos de diagnósticos secundários.

**Tabela 1** - Quantidade de diagnósticos secundários segundo o tipo de *Diagnosis Related Group* Cirúrgico e Clínico nos hospitais da pesquisa, 2012-2014.

| Número de comorbidades | Tipo | | | | Geral | |
|---|---|---|---|---|---|---|
| | Cirúrgico | | Clínico | | | |
| | n | % | n | % | n | % |
| *Nenhuma* | 43.080 | 47,4 | 12.804 | 22,0 | **55.164** | **37,9** |
| *1* | 27.134 | 29,8 | 25.497 | 46,5 | **52.631** | **36,1** |
| *2* | 7.914 | 7,9 | 5.861 | 10,7 | **13.055** | **9,0** |
| *3* | 4.631 | 5,1 | 3.826 | 7,0 | **8.457** | **5,8** |
| *4 ou +* | 8.863 | 9,8 | 7.540 | 13,8 | **16.403** | **11,2** |
| TOTAL | 90.902 | 100,0 | 54.808 | 100,0 | 145.710 | 100,0 |

**Base de Dados:** 145.710 pacientes (*Cirúrgico* → 90.902 pacientes e *Clínico* → 54.808 pacientes)

As prevalências das categorias de diagnósticos maiores (MDC), onde a categoria mais frequente do tipo Cirúrgico foi "Gravidez, Parto e Puerpério", representa 23,0%, seguida pelas "Doenças e Distúrbios do Sistema Músculo-esquelético e Tecido Conjuntivo" (15,1%) e "Doenças e Distúrbios do Sistema Digestivo" (12,0%). Esses três grupos somam cerca de 50,1% dos casos. A categoria MDC de "Doenças e Distúrbios do Sistema Respiratório" é a mais frequente entre as do tipo Clínico (16,1%).

A TABELA 2 apresenta a comparação entre os tipos Cirúrgico e Clínico em relação à classificação dos DRG em relação à mediana do tempo de permanência hospitalar de acordo com o critério americano de dias de internação.

A TABELA 3 mostram uma análise sintetizada dos resultados, considerando o tipo de DRG e percentis.

O GRÁFICO 1 mostra a análise de correlação entre o peso da complexidade assistencial e mediana do tempo de permanência hospitalar (em dias) observado em cada DRG. Não houve diferenças entre a utilização da média ou da mediana.

**Tabela 2** - Comparação entre os tipos Cirúrgico e Clínico de acordo com a classificação dos DRG em relação à mediana do tempo de permanência hospitalar segundo o critério americano, hospitais estudados, 2012-2014.

| Permanência Hospitalar - Critério americano - | Tipo | | | | |
|---|---|---|---|---|---|
| | Cirúrgico | | Clínico | | |
| | n | % | n | % | TOTAL |
| ≤ *Mediana (P$_{50}$)* | 95 | 50,8 | 48 | 20,3 | **143** |
| = *Mediana (P$_{50}$)* | 28 | 15,0 | 38 | 16,0 | **66** |
| > *Mediana (P$_{50}$)* | 64 | 34,2 | 151 | 63,7 | **215** |
| TOTAL | 187 | 100,0 | 237 | 100,0 | 424 |

**Base de Dados:** 424 DRG (*Cirúrgico* → 187 DRG e *Clínico* → 237 DRG)
**NOTA:** $p < 0,001$ (Teste *Qui-quadrado*)
  O.R. → 3,4

**Tabela 3** - Classificação dos DRG quanto à mediana do tempo de permanência hospitalar em relação aos percentis do tempo de internação americano, hospitais estudados, 2012-2014.

| Classificação dos DRG | Tipo | | | | Geral | |
|---|---|---|---|---|---|---|
| | Cirúrgico | | Clínico | | | |
| | N | % | n | % | n | % |
| **Percentis** | | | | | | |
| < $P_{10}$ | 21 | 11,2 | 7 | 3,0 | **28** | **6,6** |
| $P_{10}$ | 37 | 19,8 | 6 | 2,5 | **43** | **10,1** |
| $P_{10-25}$ | 11 | 5,9 | 3 | 1,3 | **14** | **3,3** |
| $P_{25}$ | 12 | 6,4 | 18 | 7,6 | **30** | **7,1** |
| $P_{25-50}$ | 14 | 7,5 | 14 | 5,9 | **28** | **6,6** |
| $P_{50}$ | 28 | 15,0 | 38 | 16,0 | **66** | **15,6** |
| $P_{50-75}$ | 6 | 3,2 | 9 | 3,8 | **15** | **3,6** |
| $P_{75}$ | 30 | 16,0 | 85 | 35,8 | **115** | **27,1** |
| $P_{75-90}$ | 7 | 3,7 | 12 | 5,1 | **19** | **4,5** |
| $P_{90}$ | 16 | 8,6 | 38 | 16,0 | **54** | **12,7** |
| > $P_{90}$ | 5 | 2,7 | 7 | 3,0 | **12** | **2,8** |
| **TOTAL** | **187** | **100,0** | **237** | **100,0** | **424** | **100,0** |

**Base de Dados**: 424 DRG (Cirúrgico → 187 DRG e Clínico → 237 DRG)

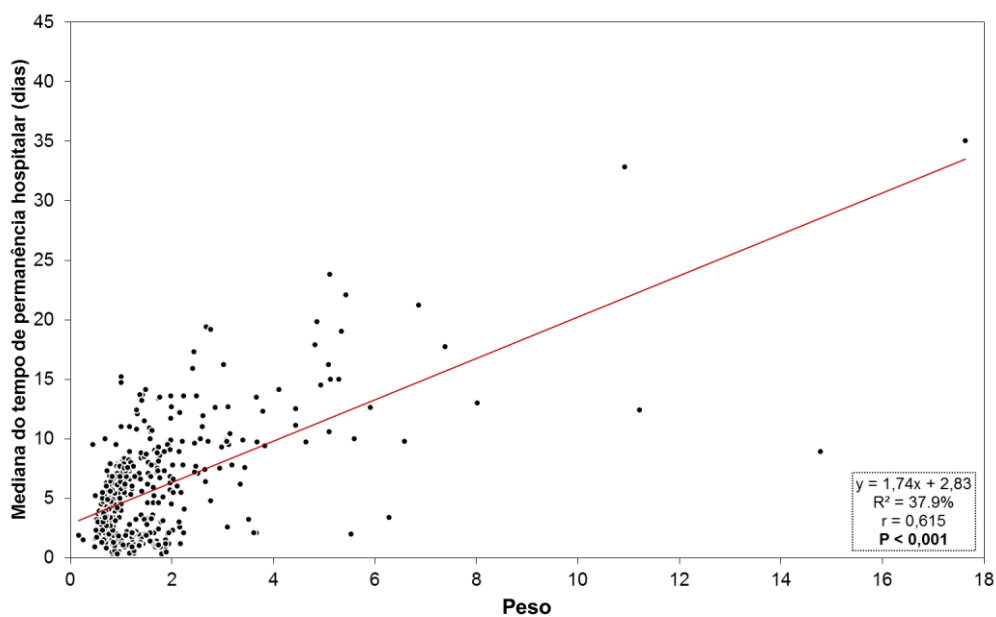

**Gráfico 1**: Análise de correlação entre o peso da complexidade assistencial e a mediana do tempo de permanência hospitalar (em dias) observado em cada DRG

# Discussão

Uma das características dos serviços de saúde é a grande heterogeneidade das informações entre prestadores e consumidores, e particularmente entre gestores. Essa dissonância torna difícil estimar o desempenho dos prestadores, afetando não apenas a escolha do paciente, mas também as decisões governamentais.

O *Diagnosis Related Groups* (DRG) é um sistema de classificação de pacientes, descrito em 1977 nos EUA, que reflete a média da relação do tipo de paciente tratado no hospital e o custo desse hospital.

Entre as versões, o IR DRG (*International Refined* – DRG) foi desenvolvido a partir de casuísticas de vários países, o que lhe confere mais representatividade, além de utilizar o CID-10 e agregar três níveis de gravidade [7].

Diversas adaptações foram desenvolvidas em diferentes países. O MS-DRG foi introduzido em 2008, na versão 25. O presente estudo, utilizando a versão 31.0 (MS-DRG), avaliou 145.710 internações de hospitais distribuídos nos estados brasileiros de Minas Gerais e Goiás.

Noronha *et al.* excluíram os casos que apresentavam tempo de permanência 2,5 desvios-padrão acima ou abaixo da média do DRG a que pertenciam, justificando que esses influenciavam os resultados estatísticos [5]. Afim de evitar vieses de seleção, em nosso estudo, optamos por manter todas as entradas originais do banco de dados e efetuar análises estatísticas adequadas à distribuição de cada variável.

## Perfil dos pacientes estudados

Poucos estudos relacionados ao DRG descrevem as características da população incluída [8,9].

Em nosso estudo, a distribuição de pacientes quanto ao sexo mostrou que as internações femininas representam dois terços do total (63,9% x 36,1%) com média de idade de 42,8 anos. As internações relacionadas a "Gravidez, Parto e Puerpério" ocupam o primeiro lugar entre os DRG Cirúrgicos (23%), agrupados no MDC 4, representando 15,5% do total. Assim, os partos espontâneos e as cesarianas, são os CID mais frequentes e podem explicar parte da maior prevalência de mulheres, com implicações na redução da média de idade.

Os diagnósticos secundários somam 78,0% das internações por DRG clínicos, sendo 20,8% com três ou mais diagnósticos, enquanto 47,4% dos DRG cirúrgicos não tiveram diagnóstico secundário. O diagnóstico secundário ao mesmo tempo que mostra esmero no exame do paciente, indica maior número de comorbidades e, por conseguinte, gera maior complexidade, o que pode determinar maior custo.

## Produtividade dos leitos

Embora a abordagem DRG represente um avanço no cálculo de pagamentos, o sistema de reembolso DRG, usado nos EUA na década de 80, não impediu a transferência de custos para os pacientes de planos privados, como também não envolveu médicos nos esforços para melhorar a eficiência de custos [10].

O DRG foi avaliado na Suíça como base para o reembolso dos prestadores. Houve ganho na prática de atendimento ambulatorial, no sentido de cooperação entre o hospital e o médico, porém ocorreu aumento do número de reinternações. Já a redução do tempo de internação, que poderia trazer reflexos positivos na produtividade, não foi relacionada à implantação do DRG. Não foram observados outros efeitos negativos [9].

O G-DRG (*Germany Diagnosis Related Group*) adotado na Alemanha em 2004, com base no AR-DRG (*Australian Refined* DRG), calcula anualmente os custos e seus pesos a partir de dados do ano anterior, sendo que em 2009 utilizou 263 (16%) hospitais da rede. Registrou-se alta qualidade na alocação de recursos hospitalares, porém limitações na delimitação do custo total de gerenciamento e do custeio individualizado do paciente. Segundo os autores, o Sistema apresentou, ainda, viés de representatividade ao adotar um modelo nacional único [11].

Em hospitais portugueses avaliaram-se eficiência e qualidade de entidades públicas e privadas. Os autores construíram um indicador agregado de avaliação da qualidade, composto por 12 indicadores de processo e cinco indicadores de resultado, e utilizaram o inverso do índice de *Casemix*, para ajustar os indicadores de gestão. Concluíram que a homogeneização obtida permitiu a comparação entre os hospitais, sem no entanto, alçar a validação da metodologia [12].

Em nosso estudo, quando se comparam os percentis dos hospitais americanos à mediana do tempo de permanência hospitalar, para cada um dos DRG com no mínimo 20 internações, observam-se duas situações distintas. Na primeira, tendo como exemplo o DRG de número 25, em que o resultado não mostra diferença estatisticamente significativa (p > 0,05) das medidas do tempo de permanência hospitalar, para este DRG, em relação ao percentil 75, isto é, o grupo de pacientes avaliado apresenta uma mediana semelhante ao percentil 75 do critério americano. Portanto, pode-se concluir que o tempo de permanência hospitalar dos hospitais avaliados está estatisticamente elevado se tomar como ponto de referência a mediana (P50) do critério americano. Isso equivale à baixa produtividade. Na outra situação, destacam-se como exemplo o DRG de número 3, em que a mediana observada difere significativamente (p < 0,05) nos 5 percentis do critério americano e, no DRG 3, a mediana encontra-se entre os percentis 50 e 75 (P50-75), portanto, está estatisticamente acima do percentil 50 (P50). Esta última conclusão baseia-se na inversão da distribuição dos sinais "+" e "-" da comparação com os valores pré-estabelecidos nos percentis americanos, obtidos na aplicação do teste de sinal [13].

Comparando-se os hospitais em estudo com os hospitais americanos, a observação dos DRG com mediana igual ou menor que o percentil 10, registra 31% de cirúrgicos e 5,5% de clínicos, o que sinaliza melhor eficiência do atendimento de pacientes vinculados aos DRG de casos cirúrgicos.

Essa percepção é reforçada nos resultados, que mostram associação significativa (p<0,001; O.R. 3,4) entre o tipo e a classificação dos DRG em relação a mediana do tempo de permanência hospitalar obtida nos hospitais americanos. A proporção de DRG do tipo cirúrgico (65,8%) é maior do que a encontrada nos DRG clínicos (36,3%), resultando daí que o tipo cirúrgico tem 3,4 vezes mais chance de apresentar um tempo de permanência hospitalar mediano abaixo ou igual ao percentil 50 americano, do que um DRG do tipo clínico.

A mediana do tempo de permanência hospitalar, avaliada no presente estudo, é comparável à dos hospitais americanos em 15,6% dos DRG, sendo que se encontra abaixo em 33,7% e acima em 50,7%. No entanto, quando se avaliam os subgrupos Clínicos e Cirúrgicos, observa-se que a permanência nos hospitais brasileiros é maior para os DRG clínicos (63,7%), contrapondo-se com 34,2% dos cirúrgicos. Considerando que maior permanência no leito implica em maior custo, o tempo de permanência será inversamente proporcional a produtividade.

Assim, pode-se inferir que a produtividade do leito clínico dos hospitais brasileiros é menor que a americana, enquanto que a dos cirúrgicos é igual ou maior (65,8%). A alta precoce de casos cirúrgicos, praticada nos hospitais avaliados, pode explicar esse resultado. Já, diversos fatores, como permanência no hospital por falta de transporte ou aguardando acompanhantes e "internações sociais", podem explicar a baixa produtividade dos casos clínicos. No estudo de Hendy *et al.*, os atrasos de alta, decorrentes de demora na finalização da terapia e motivos sociais, foram responsáveis por 26,8% dos custos [14]. Em outro estudo, 67% dos atrasos ocorreram com pacientes clinicamente estáveis e em condições de alta. Entre as causas médicas, 54% foram devido a atrasos em procedimentos, 21% aguardando a realização de exames e 10% a interpretação dos mesmos. Entre as causas não médicas destacam-se a dificuldade de contatar familiares e transferir pacientes para casas de repouso, além de problemas relacionados ao transporte [15]. Em nosso estudo, não avaliamos esses motivos que resultaram na melhor produtividade de pacientes cirúrgicos e no maior custo dos pacientes clínicos, em que pese esses serem os principais responsáveis pela baixa produtividade em nosso meio, devendo ser objeto de atenção na otimização do sistema de saúde.

### Correlações do peso da complexidade assistencial

A relação entre a complexidade produtiva de cada produto assistencial DRG, foi medida pelo peso do produto na composição do *Casemix* hospitalar e as respectivas variações de produtividade. Tal complexidade, relacionada ao consumo de recursos, tem sido utilizada como um dos critérios para a alocação de recursos financeiros a hospitais [16,17,18].

Muitos países calculam os pesos relativos do DRG ou adaptam de outros países, como o fazem Portugal e Irlanda. Já, Inglaterra, França, Holanda e Espanha utilizam valores monetários dos custos, obtidos sem cálculo de peso. Por outro lado, Áustria e Polônia são os únicos que expressam o peso do DRG como escore. A diferença é que o escore não expressa valor monetário, mas um número de pontos. Em contraste com o peso relativo, o escore não guarda relação com a média de custo de tratamento de cada DRG do país. As nações europeias utilizam, ainda, diferentes critérios de conversão monetária, que podem variar dentro do mesmo país [19].

Em estudo no Canadá, os programas de garantia da qualidade para unidades de reabilitação de internações por AVC na província de Quebec, tiveram como principal objetivo uma redução anual de tempo médio de permanência (LOS – *lenght of stay)* para controlar progressivamente os custos de saúde e, possivelmente, aumentar o número de pacientes atendidos em um ano fiscal. Esta prática pode levar a desigualdades entre provedores de reabilitação, uma vez que existem potenciais incentivos para selecionar preferencialmente indivíduos com deficiências menos complexas e deficiências combinadas com a maioria dos ambientes sociais e físicos favoráveis, a fim de alcançar os objetivos visados. Perniciosamente, a maioria dos indivíduos com níveis de imparidade e deficiência grave pode ser confrontado com acessibilidade limitada, ou mesmo ilegibilidade, a programas de reabilitação neurológica. Além disso, instalações de reabilitação podem cumprir os objetivos anuais pela aplicação de uma estratégia de mudança de custo, que consiste principalmente de aumentar as taxas de referência para outros provedores de reabilitação (por exemplo, provedores de *home care*). Portanto, a possibilidade de usar o sistema de classificação e implementação de grupo de *Casemix* vem despertando interesse considerável [8].

Embora tenha utilizado o *Casemix* como uma medida de complexidade da enfermidade do paciente e do tratamento associado, seu valor não foi informado no estudo realizado no Chile [20].
Na avaliação de pacientes internados com doenças respiratórias, que representaram um quarto das internações, Cots e colaboradores utilizaram o peso dos DRG respiratórios (1,94) em relação ao peso do *Casemix* dos pacientes clínicos (1,77) para apontar a maior complexidade dessas doenças [21].

Considera-se o tempo em dias de permanência no leito hospitalar, como o principal fator que está ligado diretamente ao custo e desempenho do produto assistencial. Entende-se por complexidade assistencial, em especial, as condições biológicas, gravidade da doença e o manejo clínico. No presente estudo, o peso da complexidade assistencial de cada produto apresentou correlação positiva com o tempo mediano de permanência hospitalar. A diferença entre o grau de correlação dos DRG clínicos ($p=0,001$; $R^2=28,2$) em relação aos cirúrgicos ($p=0,001$; $R^2=49,6$) não explica os achados descritos anteriormente, que mostraram produtividade dos leitos clínicos dos hospitais brasileiros menor que a dos hospitais americanos em 63,7% dos DRG, enquanto que a dos cirúrgicos foi igual ou maior em 65,8%.

Na categoria de DRG cirúrgicos a correlação encontrada ($p=0,001$; $R^2=49,6$), mostra que a complexidade explica cerca de metade dos fatores que impactam no tempo de ocupação do leito, mas faz procurar motivos que expliquem a aproximação da produtividade do leito cirúrgico dos hospitais estudados à dos americanos.

De outro lado, nos DRG clínicos a parte não correlacionada com a complexidade da assistência, a maior parte, representando cerca de 70%, decorre de outros fatores, não biológicos, inerentes ao sistema de saúde. Assim, o índice de risco de complexidade é um determinante desta relação, mas não o mais importante, e sugere que há outros fatores impactando e alongando o tempo de ocupação dos leitos hospitalares. Esta é a grande janela de oportunidade de melhoria na reestruturação dos processos e custos hospitalares, sem impactar na assistência aos pacientes.

## Conclusões

A população estudada é predominantemente feminina com média de idade de 42,8 anos. A produtividade dos leitos clínicos dos hospitais estudados é menor que a dos hospitais americanos em 63,7% dos DRG, enquanto que a dos cirúrgicos é igual ou maior em 65,8%.

Parte dos determinantes da longa permanência nos leitos de hospitais brasileiros podem ser explicados por causas não biológicas. Neste sentido faz-se necessário uma mudança no paradigma da gestão com foco nos processos hospitalares. Fatores diferentes da complexidade assistencial influenciam mais que 70% da correlação de custos dos leitos clínicos, mostrando desta forma uma grande oportunidade para melhoria dos sistemas de saúde.

## Referências